\documentstyle[12pt,epsf,fleqn]{elsart}

\def \be {\begin{equation}}
\def \ee {\end{equation}}

\begin{document}

\begin{frontmatter}
\title{GENIUS-TF: a test facility for the GENIUS project}
\noindent
\author{L. Baudis$^{*}$, A. Dietz, G. Heusser, B. Majorovits,}
\author{H. Strecker and H. V. Klapdor--Kleingrothaus$^{**}$}
\address{Max--Planck--Institut f\"ur Kernphysik, Heidelberg, Germany}

\begin{abstract}
GENIUS is a proposal for a large scale detector of rare events.
As a first step of the experiment, a small test version, the GENIUS
test facility, 
will be build up at the Laboratorio Nazionale del Gran Sasso (LNGS).
With about 40\,kg of natural Ge detectors operated in liquid nitrogen,
GENIUS-TF could exclude (or directly confirm) the DAMA annual modulation
signature within about two years of measurement.
\end{abstract}
\end{frontmatter}

\section{Introduction}

GENIUS (GErmanium in liquid NItrogen Underground Setup) is a proposal
for operating a large amount of 'naked' Ge detectors in liquid
nitrogen to search for rare events such as WIMP-nucleus scattering,
neutrinoless double beta decay and solar neutrino interactions, 
with a much increased sensitivity relative to existing experiments
 \cite{ringb,nim_genius,prop_genius}.
By removing (almost) all materials from the immediate vicinity  of 
the Ge-crystals, their absolute background can be considerably
decreased with respect to conventionally operated detectors. 
The liquid nitrogen acts both as a cooling medium and as a shield
against external radioactivity. 
The proposed scale of the experiment 
is a nitrogen tank of about 12\,m diameter and 12\,m height 
with 100\,kg of natural Ge and 1 ton of enriched $^{76}$Ge in 
the dark matter and double beta decay versions, respectively,
suspended in its center.

To cover large parts of the MSSM parameter space, relevant for the
detection of neutralinos as  dark matter candidates \cite{jkg96,bedny1,bedny2}, 
a maximum background
level of 10$^{-2}$ counts/(kg\,y\,keV) in the energy region below 
50\,keV has  to be achieved. In the double beta decay 
region (Q-value = 2038.56\,keV) a background of 0.3~events/(t\,y\,keV)
is needed in order to test the effective Majorana neutrino mass down
to 0.01\,eV (90\%C.L.).
This implies a very large background reduction in comparison to our recent best results 
\cite{prd-hdmo,prl-hdmo}  with the Heidelberg--Moscow experiment.

The focus of this paper is to present a small scale version
experiment, the GENIUS test facility (GENIUS-TF), which will be built up at the  
Laboratorio Nazionale del Gran Sasso (LNGS).
It is designed to test experimentally the feasibility of GENIUS. 
Up to 40\,kg of Ge detectors 
will be operated directly in liquid nitrogen, the overall dimension of 
the experiment not exceeding 2\,m$\times$2\,m$\times$2\,m.
As a side effect, it will improve limits on WIMP-nucleon cross sections with respect  
to our results  with the Heidelberg--Moscow and HDMS experiments \cite{prd-hdmo,prd-hdms}.
The relatively large mass of Ge compared to existing
experiments would permit to search directly for a WIMP signature in
form of the  predicted \cite{freese} seasonal modulation of the event rate. 

\section{The GENIUS Test Facility}

\begin{figure}
\epsfysize=90mm\centerline{\epsffile{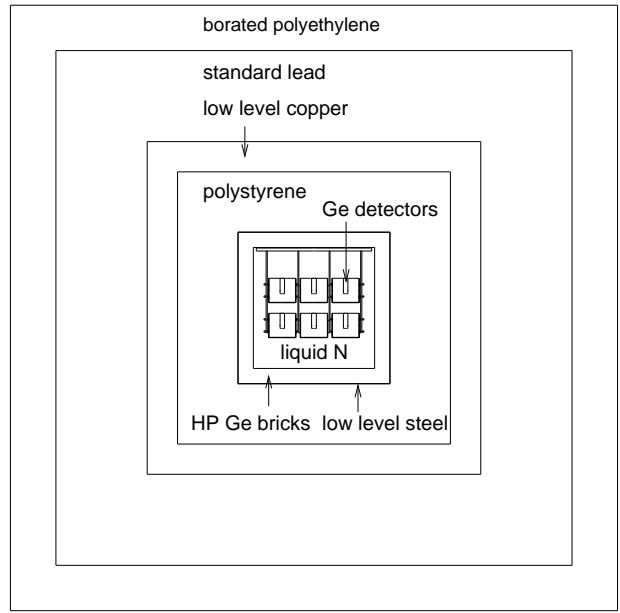}}
\caption{Schematic view of the GENIUS test facility. 
14 HPGe detectors are operated in a steel vessel 
filled with liquid nitrogen, with an inner shield of HPGe bricks,
isolated by polystyrene and surrounded by copper, lead and borated polyethylene shields. } 
\label{gtf_setup}
\end{figure}

GENIUS-TF consists of up to 14 natural Ge crystals (40\,kg) 
operated in a volume of 0.064\,m$^3$ of ultra-pure liquid nitrogen.
The liquid nitrogen is housed by a 0.5\,mm thick steel vessel inside 
a 0.9\,m$\times$0.9\,m$\times$0.9\,m  box of
polystyrene foam, with a 5\,cm thick inner shield of high purity Ge bricks 
(the basic concept is described in \cite{bargein}). Outside the foam box there
are 10\,cm of low-level copper, 30\,cm of lead and 15\,cm of borated polyethylene as shields
against the natural radioactivity of the environment (see Fig. \ref{gtf_setup}). 
The Ge crystals are positioned in two layers (each layer of 7
detectors in two concentric circles) on a holder system made of high
molecular polyethylene. The signal and high voltage contacts of the
individual crystals are established using a minimized amount of 
ultra pure stainless steel (about 3\,g) as already demonstrated in previous
experiments \cite{nim_genius,la_diss,bela_diss}.

\section{Background considerations}

The aim of GENIUS-TF is to reach  the background
level of  2\,events/(kg\,y\,keV) in the energy region below
50\,keV (50\,keV ionization energy in germanium corresponds to about
200\,keV nuclear recoil energy).
This is one order of magnitude lower than the actual background of the 
Heidelberg-Moscow experiment and 
two orders of magnitude higher than the final goal of GENIUS.
To estimate the background contributions from the various components 
we performed detailed Monte Carlo simulations of the relevant background
sources using the geometry shown in Fig.  \ref{gtf_setup}
\footnote{in \cite{bargein} some first simulations with one detector 
in a simplified geometry had been made}. 
The simulations are based on the {\sc GEANT3.21} package 
\cite{geant} extended for nuclear decays.

The sources of background can be divided into external and internal ones.
The external background 
is generated by events originating from outside
the shields, such as photons and neutrons from the Gran Sasso
rock and by  muon interactions.
We simulated the measured photon \cite{arpesella92}, neutron \cite{arp} 
and muon \cite{macro} fluxes at LNGS. With a total of 30\,cm of lead
and 15\,cm of borated polyethylene shield, 
the contribution of the photons and neutrons are negligible.
The muons yield a count rate of  2$\times$10$^{-2}$ events/(kg\,y\,keV)
in the energy region 0-50\,keV (this, and all following 
 rates are given after computing the anti-coincidence between the 14 Ge 
 detectors). Secondary neutron induced interactions in the liquid nitrogen, as well 
as negative muon capture and inelastic muon scattering reactions 
generate only a 
negligible contribution to the overall expected background rate (for 
details see \cite{nim_genius,prop_genius,la_diss}).
 
Internal background arises from residual impurities in the liquid
nitrogen, the steel vessel, the polystyrene foam isolation, 
the Ge and  Cu shields, 
the crystal holder system, the Ge crystals
themselves and from activation of the Ge crystals and of the copper
during fabrication and transportation at the Earths surface.
The assumed intrinsic impurity levels for the simulated materials
and the resulting count rates in the low-energy region are listed in 
Table ~\ref{intr-assumpt}.

The values assumed for the $^{238}$U and $^{232}$Th contamination of the 
liquid nitrogen are 1000 times higher than already measured
by BOREXINO \cite{borexino} for their liquid scintillator. 
The $^{222}$Rn contamination of freshly
produced liquid nitrogen was recently measured to be 325\,$\mu$Bq/m$^3$ 
\cite{rau}. No additional assumptions were made.
The U/Th, $^{40}$K and $^{60}$Co contamination for the steel are taken 
from a recent measurement in the Heidelberg low level lab \cite{heusser-privat}.
The intrinsic impurity levels in Ge crystals are 
conservative upper limits from measurements with the detectors of the
Heidelberg--Moscow experiment. We assumed a 100 times higher
contamination level for the HPGe bricks used as inner shield. 
The contamination level of polystyrene was measured with a natural Ge
detector in appropriate shielding at LNGS \cite{bela_diss}. 
However, no material selection or special handling were applied and a
higher purity can certainly be reached. 
The $^{238}$U, $^{232}$Th and $^{40}$K contamination values, as well 
as the cosmogenic activation  of the Cu shield were taken from a
former measurement  with the Ge detectors of
the Heidelberg-Moscow experiment \cite{hdmo-old}. In the
Heidelberg-Moscow experiment we also measure activities of the
anthropogenic isotopes $^{125}$Sb, $^{207}$Bi, $^{134}$Cs, $^{137}$Cs.
Although these impurities could be found anywhere in the
experimental setup, we assume that they are located in the copper shield.
The respective activities are taken from \cite{maier_diss}.
The assumed values for polyethylene were  reached by the SNO experiment
\cite{sno} for an acrylic material. Though a 10-100 times higher 
contamination level would also be acceptable, such a crystal support
system still has to be developed.

\begin{figure}
\epsfysize=83mm\centerline{\epsffile{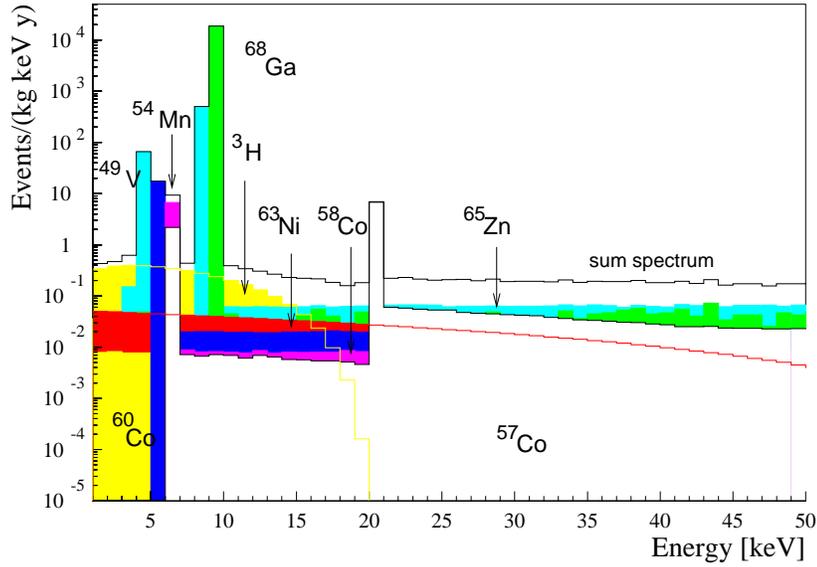}}
\caption{Background originating from cosmic activation of the Ge
  crystals at sea level with 30 days exposure and 1\,year deactivation 
  below ground.} 
\label{cosmo}
\end{figure}

We have estimated the cosmogenic production rates of radioisotopes 
in the Ge--crystals with the $\Sigma$ program \cite{JensB}.
Assuming an unshielded production and transportation time of 30 days at sea level
for the Ge--detectors, and a deactivation time of one year,
we obtain the radioisotope concentrations listed in Table
\ref{ge_cosmo} (for $^{68}$Ge the saturation activity is assumed; the 
value for $^{3}$H is taken from \cite{avignone}).
All other produced radionuclides have much smaller activities due to their shorter
half lives. 
The count rate between 5\,keV and 11\,keV is dominated by X--rays from 
the decays of the various isotopes (see Table \ref{ge_cosmo}).        
However, if the  energy threshold of the Ge detectors will be as low 
as 0.7\,keV (as stated by the manufacturer)  GENIUS-TF will nonetheless 
be sensitive to low WIMP masses.
The sum of all contributions from the cosmogenic activation of the Ge
crystals amounts to 4$\times$10$^{-1}$ counts/(kg\,y\,keV) between
1 and 4\,keV and between 11 and 50\,keV. After the decay of those isotopes with half lives
around 1 year, this region will be dominated by contribution from
$^{3}$H and $^{63}$Ni, due to their  low Q--value (18.6\,keV and
66.95\,keV) and large half life (12.33\,yr and 100.1\,yr).
Figure \ref{cosmo} shows the sum and the single contributions from the
different isotopes.

\begin{figure}
\begin{center}
\epsfysize=85mm\centerline{\epsffile{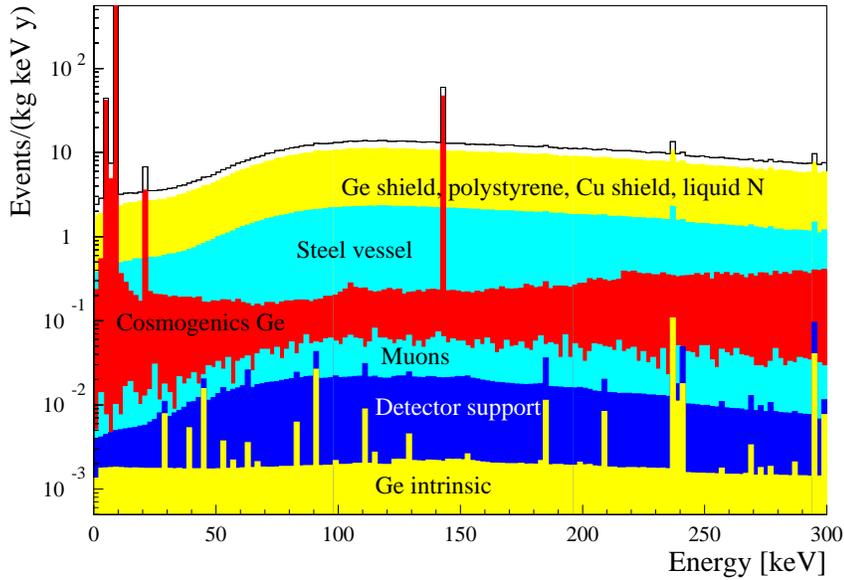}}
\end{center}
\caption{Simulated spectra of the dominant background sources for
the GENIUS test facility.
One year of storage below ground for the Ge detectors was assumed.
Shown is the low-energy region with contributions from the germanium and copper shields,
the polystyrene isolation and the detector support system,
the liquid nitrogen, cosmogenic activation of the Ge crystals and 
intrinsic impurities of the crystals and muon induced background.
The solid line represents the sum spectrum of all the simulated
components.}
\label{specall}
\end{figure}

Summing up the background contributions discussed so far,
the mean count rate  in the low energy 
region amounts to about 4 events/(kg\,y\,keV).
In Fig. \ref{specall} the spectra of individual contributions 
and the summed up total background spectrum are shown (after one
year of storage of the Ge detectors below ground). 
The low-energy spectrum is dominated by events
originating from the polystyrene foam isolations, from the copper
shield and the steel vessel.
Regarding the polystyrene, no material  selections were performed so 
far and efforts in this direction are starting to being made.
Careful selections will have to be performed also for the steel
vessel, which, in spite of its low total mass, yields a significant contribution
due to its proximity to the Ge crystals.
Lower contamination values by a factor of 5-10 than assumed in this
simulation were already reached in the past \cite{heusser-privat}.
For copper the cosmogenic activities of 
$^{54}$Mn, $^{57}$Co, $^{58}$Co, $^{60}$Co, as well as anthropogenic
activities are dominating and low exposures at the Earths 
surface as well as electro-polishing of surfaces are desirable.

\section{Goals of GENIUS-TF }

The primary goal of  GENIUS-TF is to demonstrate the feasibility of
the GENIUS project.
It has  to be shown that `naked' Ge detectors work reliably in 
liquid nitrogen over a longer period of time (at least for one year). 
Material selections have to be performed for various experimental
components  and their purity tested down to
1\,event/(kg\,y\,keV). A crystal support system, 
made of low-radioactivity polyethylene has to be developed and designed such that it
can be extended in order to house up to 40 crystals (100\,kg).
A new, modular data acquisition system and electronics have to be
developed and tested.
Besides above issues, which certainly are important, 
GENIUS-TF can have a physics program of its own.

\subsection*{WIMP Dark Matter}

With 40\,kg of natural Ge and a background of 2\,events/(kg\,y\,keV) 
in the energy region below 50\,keV, GENIUS-TF can cover the `evidence region'
in the MSSM parameter space for neutralinos as dark 
matter candidates singled out by the DAMA experiment  \cite{dama3}.
It would exclude DAMA after about one year of measurement, delivering 
an  independent test by using a different technology and 
raw data without background subtraction.

\begin{figure}
\begin{center} 
\epsfysize=90mm\centerline{\epsffile{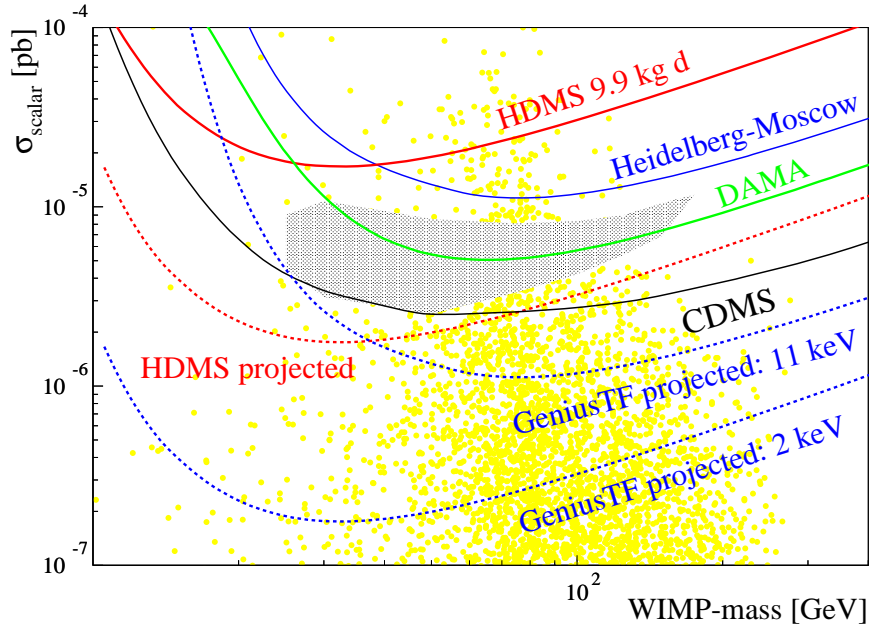}}
\caption{
WIMP-nucleon cross section limits as a function of the WIMP
  mass for spin-independent interactions. 
The solid lines are current limits of the Heidelberg-Moscow experiment 
{\protect{\cite{prd-hdmo}}}, 
the HDMS prototype {\protect{\cite{prd-hdms}}}, the DAMA experiment 
{\protect{\cite{dama}}} and the CDMS experiment 
{\protect{\cite{rick2000}}}.
The dashed curves are the expectation for HDMS
{\protect{\cite{prd-hdms}}} and for GENIUS-TF
with an energy threshold of 11\,keV and 2\,keV respectively, and a background index of 2
events/kg\,y\,keV below 50\,keV.
The  filled contour represents  the 2$\sigma$ evidence region of the DAMA
  experiment {\protect{\cite{dama3}}}.    
The experimental limits are compared to
expectations (scatter plot) for WIMP-neutralinos calculated in the
MSSM parameter space at the weak scale 
under the assumption that all superpartner masses are lower than
300 GeV - 400 GeV {\protect{\cite{vadim99}}}.}
\end{center} 
\label{limits}
\end{figure}

Figure \ref{limits} shows a comparison of existing constraints and 
future sensitivities of cold dark matter experiments, together with the 
theoretical expectations for neutralino scattering rates 
\cite{vadim99}. For GENIUS-TF, energy thresholds of 2\,keV and 11\,keV 
(worst case scenario) were assumed.
In addition to setting limits on WIMP-nucleon cross sections, GENIUS-TF 
will  be able to test the DAMA  region  by directly looking for a
seasonal modulation of the event rate and of the energy spectrum.
Depending on the background and energy threshold, an overall exposure
between 1 and 5 years are needed in order to test the DAMA region with
99.5\% C.L. (according to \cite{cebrian99}). 
For example, for an energy threshold of 2\,keV and a
background level of  4\,events/(kg\,y\,keV), 1.4\,yr of measurement
with 40\,kg of Ge are required. Even for an initial  lower mass of 20\,kg the
time scale of about 3\,yr would  still be acceptable. 

\subsection*{Neutrinoless Double Beta Decay}

Neutrinoless double beta decay provides a unique method for 
gaining information about the absolute neutrino mass scale 
and of discerning between a Majorana and a Dirac neutrino. 
The current most stringent experimental limit on the effective  
Majorana neutrino mass, $\langle {\rm m} \rangle < $0.35\,eV, comes 
from the Heidelberg-Moscow experiment \cite{hdmo_dark}.
For a significant step beyond this
limit, much higher source strengths and lower background levels are
needed, a goal which could be accomplished by the GENIUS experiment operating 
300--400 detectors made of enriched $^{76}$Ge (1 ton).

\begin{figure}[t!]
\begin{center}
\epsfysize=90mm\centerline{\epsffile{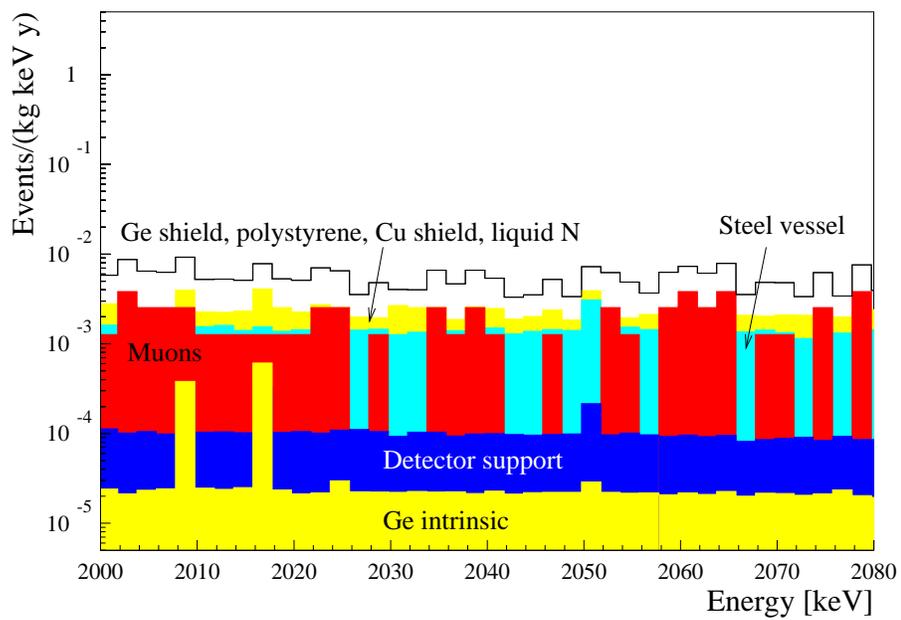}}
\end{center}
\caption{Simulated spectra of the dominant background sources for the 
enriched $^{76}$Ge detectors of the Heidelberg-Moscow experiment
in the GENIUS-TF setup. The energy region relevant for the  
search of the neutrinoless double beta decay is shown.
The solid line represents the sum spectrum of all the simulated
components.}
\label{specall-betabeta}
\end{figure}

Operating the enriched $^{76}$Ge detectors of the
Heidelberg-Moscow experiment within the GENIUS test facility could
improve existing half life limits by up to a factor of 8. 
The Heidelberg-Moscow Ge detectors have the advantage of having been
stored for several years at LNGS, so that cosmogenic
activities will not play a major role for the background.
However, in order to improve the background in the high-energy region
by a factor of 30, more stringent requirements for the purity of the
used materials have to be made (see Table \ref{intr-assumpt-bb}).
The U/Th contamination of nitrogen is 100 times less stringent than
measured by BOREXINO, the contamination of the Ge bricks 10 times
higher than the upper limit for the detectors of the  Heidelberg-Moscow experiment. 
For steel, the best measured values so far  
($^{238}$U: 0.6\,mBq/kg, $^{232}$Th: 0.2\,mBq/kg
\cite{heusser-privat}) have been assumed.
While the above components are not critical, the background is
dominated by the U/Th contaminations of the polystyrene foam and of
the copper shield. Here 50 and 10 times lower contamination levels
than measured so far have been assumed.
While for polystyrene such values could be achieved after severe material
selections, the case of copper is more subtle and it might have to be
replaced by a cleaner material, as for example low-level lead.
The muon contribution, which amounts to about
1$\times$10$^{-3}$events/kg\,yr\,keV could be further reduced by a
factor 10 with a muon veto with 90\% efficiency.
Fig. \ref{specall-betabeta} shows the individual contributions and 
the sum spectrum from 2000 to 2080\,keV.
The background of 6$\times$10$^{-3}$events/kg\,yr\,keV 
is about a factor of 30 lower than the (raw) background of the
Heidelberg-Moscow experiment in the energy region between
2000-2080\,keV \cite{la_diss} and can be further reduced by a factor
of 3 with pulse shape analysis \cite{prl-hdmo}. 
It would allow to reach a half life limit of about 1.6$\times$10$^{26}$\,yr  
and thus to test the effective Majorana
neutrino mass down to 0.1\,eV with 90\% C.L. after 6 years of measurement.
 Although 6 years seem long, the time scale is short 
compared to those of other experiments proposed to improve 
the Heidelberg-Moscow mass limit by a significant amount, 
such as GENIUS \cite{prop_genius}, EXO \cite{exo} or CUORE \cite{cuore}.

\section{Summary and Outlook}

We have presented a test facility for the GENIUS experiment, GENIUS-TF,
which is approved and is going to be installed at LNGS in the course 
of the year 2001. 
We have estimated the expected background from the various 
experimental components in a detailed Monte Carlo simulation
based on {\sc GEANT3.21}. 
A background reduction by a factor of 10 compared to the
Heidelberg-Moscow experiment \cite{prd-hdmo} seems feasible 
even with 30 days exposure of the Ge crystals at the Earths surface.
The dominating sources of background arise from the U/Th contamination
of the polystyrene foam isolation, from the steel vessel and from the
cosmogenic activation and anthropogenic contamination  
of the copper shield. While for  polystyrene and steel material 
selections will be pursued, low exposure times and special treatment
of the copper surfaces will be essential.

Besides offering an environment to test different solutions to be 
adopted in the GENIUS experiment, GENIUS-TF will bring its own 
contribution to the fields of direct WIMP detections and neutrinoless 
double beta decay search. 
It will allow to improve current limits on WIMP-nucleon cross sections 
and thus to test the DAMA evidence region \cite{dama3} within about
one year of measurement.
Moreover, with 40\,kg of WIMP target material and an energy threshold
of 1\,keV (11\,keV in the worst case), an eventual WIMP signature
could be seen (or excluded) directly within 1\,year (5\,years) of
measurement. No other planned experiment for the near future and
using a different technique than DAMA can achieve this goal.
The operation of the Heidelberg-Moscow enriched  $^{76}$Ge detectors (about
11\,kg of active mass) in the same facility would allow to test the
effective Majorana neutrino mass down to 0.1\,eV with 90\% C.L. within 
6 years of measurement.

$^{*}$ email: laura.baudis@mpi-hd.mpg.de\\
$^{**}$ spokesman of the GENIUS collab.,  email: klapdor@daniel.mpi-hd.mpg.de

\newpage

\begin{table}
\caption{Assumed intrinsic impurity levels and resulting count rates for the simulated 
background components in the energy region relevant for dark matter detection.}
\begin{center}
\renewcommand{\arraystretch}{1.4}
\setlength\tabcolsep{7pt}
\begin{tabular}{llll}
\hline\noalign{\smallskip}
Source & Radionuclide & Purity & Count rate (0-50\,keV) \\
       &              &        &  [events/(kg\,y\,keV)] \\
\hline\noalign{\smallskip}
Ge crystals  &  $^{238}$U  & 1.8$\times$10$^{-15}$g/g &  1.2$\times$10$^{-3}$ \\
             &  $^{232}$Th & 5.7$\times$10$^{-15}$g/g &  0.5$\times$10$^{-3}$ \\
\hline  \noalign{\smallskip}   
Holder system  & U/Th, K   & 1$\times$10$^{-12}$, 1$\times$10$^{-9}$g/g & 1$\times$10$^{-2}$ \\
\hline\noalign{\smallskip}
Nitrogen   & $^{238}$U  & 3.5$\times$10$^{-13}$g/g  &  3$\times$10$^{-2}$ \\
           & $^{232}$Th & 4.4$\times$10$^{-13}$g/g  &  2$\times$10$^{-2}$ \\
           & $^{40}$K   & 1$\times$10$^{-11}$g/g    &  3$\times$10$^{-3}$ \\
           & $^{222}$Rn &   325\,$\mu$Bq/m$^3$        &  4$\times$10$^{-3}$ \\ 
\hline\noalign{\smallskip}
Steel      & $^{238}$U  & 3\,mBq/kg  &   3$\times$10$^{-1}$ \\
           & $^{232}$Th & 4\,mBq/kg  &   4.5$\times$10$^{-1}$ \\
           & $^{40}$K   & 2\,mBq/kg  &   1.5$\times$10$^{-2}$ \\
           & $^{60}$Co &  2\,mBq/kg  &   3$\times$10$^{-1}$ \\ 
\hline  \noalign{\smallskip}   
Ge shield  &  $^{238}$U  & 1.8$\times$10$^{-13}$g/g &  4.5$\times$10$^{-2}$ \\
           &  $^{232}$Th & 5.7$\times$10$^{-13}$g/g &  6$\times$10$^{-2}$ \\
           &  $^{40}$K   & 1$\times$10$^{-11}$g/g      &  4.5$\times$10$^{-4}$ \\
\hline  \noalign{\smallskip}   
Polystyrene shield  &  $^{238}$U  & 1.7$\times$10$^{-10}$g/g &  1.5$\times$10$^{-1}$ \\
                   &  $^{232}$Th & 1.8$\times$10$^{-9}$g/g &  6$\times$10$^{-1}$ \\
                   &  $^{40}$K   & 2.6$\times$10$^{-7}$g/g      &  4$\times$10$^{-2}$ \\
\hline  \noalign{\smallskip}   
Cu shield  &  $^{238}$U  & 5.4$\times$10$^{-12}$g/g &  2.5$\times$10$^{-1}$ \\
           &  $^{232}$Th & 3.0$\times$10$^{-12}$g/g &  6.5$\times$10$^{-2}$ \\
           &  $^{40}$K   & 4.5$\times$10$^{-10}$g/g      &  6$\times$10$^{-3}$ \\
cosmogenics           &  $^{54}$Mn, $^{57}$Co, $^{58}$Co, $^{60}$Co &
              23,30,50,70\,$\mu$Bq/kg &  8$\times$10$^{-1}$ \\
anthropogenics           &  $^{125}$Sb, $^{207}$Bi, $^{134}$Cs, $^{137}$Cs &
              50,8,150,11\,$\mu$Bq/kg & 7.5$\times$10$^{-1}$ \\
\hline
\end{tabular}
\end{center}
\label{intr-assumpt}
\end{table}

\newpage

\begin{table}
\caption{Cosmogenically produced isotopes in the Ge crystals for an
  exposure at sea level of 30 days and a subsequent deep underground
  storage of 1\,year (for $^{68}$Ge the saturation activity was assumed)}
\begin{center}
\renewcommand{\arraystretch}{1.4}
\setlength\tabcolsep{5pt}
\begin{tabular}{llll}
\hline\noalign{\smallskip}
Isotope & Decay, T$_{1/2}$ & Energy deposition in the crystal [keV]& 
A [$\mu$Bq kg$^{-1}$]\\
\hline\noalign{\smallskip}
$^{3}$H   &   $\beta ^-$, 12.33\,yr    & E$_{\beta^-}$= 18.6\,keV  & 0.12\\
$^{49}$V   &  EC, 330\,d    & E$_K$(Ti)= 5, no $\gamma$ & 2.4\\
$^{54}$Mn  &  EC+$\beta ^+$, 312.3\,d  & E$_{\gamma}$= 840.8, E$_K$(Cr)= 5.4    & 3.1\\
$^{55}$Fe  &  EC, 2.73\,yr   & E$_K$(Mn)= 6, no $\gamma$    & 1.6\\
$^{57}$Co  &  EC, 271.8\,d  & E$_{\gamma}$= 20.81,142.8, E$_K$(Fe)= 6.4  & 3.5\\
$^{58}$Co  &  EC+$\beta ^+$, 70.9\,d   &  E$_{\gamma}$= 817.2, E$_K$(Fe)= 6.4  & 1.5  \\
$^{60}$Co  &  $\beta ^-$, 5.27\,yr & E$_{\beta^-}$= 318, E$_{\gamma}$=1173.2,1332.5  &0.7\\
$^{63}$Ni & $\beta^-$, 100.1\,yr & E$_{\beta^-}$= 66.95, no $\gamma$ & 0.04\\
$^{65}$Zn & EC+$\beta ^+$, 244.3\,d & E$_{\gamma}$=1124.4, E$_K$(Cu)=8-9  & 27\\
$^{68}$Ge & EC, 270.8\,d & E$_K$(Ga)=10.37 &676\\
$^{68}$Ga & EC+$\beta ^+$, 67.6\,m & Q-value=2921.1 &676\\
\hline
\end{tabular}
\end{center}
\label{ge_cosmo}
\end{table}

\newpage

\begin{table}
\caption{Assumed intrinsic impurity levels  and resulting count rates for the simulated 
background components in the energy region relevant for neutrinoless
double beta decay in $^{76}$Ge.}
\begin{center}
\renewcommand{\arraystretch}{1.4}
\setlength\tabcolsep{7pt}
\begin{tabular}{llll}
\hline\noalign{\smallskip}
Source & Radionuclide & Purity & Count rate (2000-2080\,keV) \\
       &              &        &  [events/(kg\,y\,keV)] \\
\hline\noalign{\smallskip}
Ge crystals  &  $^{238}$U  & 1.8$\times$10$^{-15}$g/g &  2$\times$10$^{-5}$ \\
             &  $^{232}$Th & 5.7$\times$10$^{-15}$g/g &  1$\times$10$^{-5}$ \\
\hline  \noalign{\smallskip}   
Holder system  & U/Th   & 1$\times$10$^{-12}$ & 1$\times$10$^{-5}$ \\
\hline\noalign{\smallskip}
Nitrogen   & $^{238}$U  & 3.5$\times$10$^{-14}$g/g  &  7$\times$10$^{-5}$ \\
           & $^{232}$Th & 4.4$\times$10$^{-14}$g/g  &  4$\times$10$^{-5}$ \\
           & $^{222}$Rn &   325\,$\mu$Bq/m$^3$        &  6$\times$10$^{-5}$ \\ 
\hline\noalign{\smallskip}
Steel      & $^{238}$U  & 0.6\,mBq/kg  &   6$\times$10$^{-4}$ \\
           & $^{232}$Th & 0.2\,mBq/kg  &   8$\times$10$^{-4}$ \\
           & $^{60}$Co &  2\,mBq/kg    &   1$\times$10$^{-5}$ \\ 
\hline  \noalign{\smallskip}   
Ge shield  &  $^{238}$U  & 1.8$\times$10$^{-14}$g/g &  4$\times$10$^{-5}$ \\
           &  $^{232}$Th & 5.7$\times$10$^{-14}$g/g &  2$\times$10$^{-4}$ \\
\hline  \noalign{\smallskip}   
Polystyrene shield  &  $^{238}$U  & 3.4$\times$10$^{-12}$g/g &  5$\times$10$^{-5}$ \\
                   &  $^{232}$Th & 3.6$\times$10$^{-11}$g/g  &  6$\times$10$^{-4}$ \\
\hline  \noalign{\smallskip}   
Cu shield  &  $^{238}$U  & 5.4$\times$10$^{-13}$g/g &  4$\times$10$^{-4}$ \\
           &  $^{232}$Th & 3.0$\times$10$^{-13}$g/g &  4$\times$10$^{-4}$ \\
\hline
\end{tabular}
\end{center}
\label{intr-assumpt-bb}
\end{table}

\end{document}